\documentstyle[12pt]{article}
\begin{document}
\setcounter{page}{0}
\begin{flushright}
{CERN--TH/95--47\\
DOE/ER/40427--04--N95}
\end{flushright}
\vskip0.5cm
\begin{center}

{\large{\bf THE PROTON SPIN PUZZLE AND\\
\vskip0.2cm
 DEPOLARIZATION IN
$\bar{p}p \to \bar{\Lambda}\Lambda$}}

\vskip1cm

Mary Alberg$^{a,b,}$\footnote{E-mail: alberg@phys.washington.edu},
John Ellis$^{c,}$\footnote{E-mail: johne@cernvm.cern.ch}
and Dmitri Kharzeev$^{c,d,}$\footnote{E-mail: kharzeev@vxcern.cern.ch}\\

\vskip0.8cm
{\it $^a)$ Department of Physics\\
Seattle University\\
Seattle, WA 98122, USA}
\medskip

{\it $^b)$ Department of Physics\\
University of Washington\\
Seattle, WA 98195, USA}
\medskip

{\it $ ^c)$ Theory Division\\
CERN\\
Geneva, Switzerland}
\medskip

{\it $^d)$ Physics Department\\
University of Bielefeld\\
33615 Bielefeld, Germany}
\end{center}
\medskip
\begin{abstract}
We point out that the measurement of target spin depolarization $D_{nn}$ in
the $\bar{p}p\to\bar{\Lambda}\Lambda$ reaction may test dynamical
mechanisms invoked to explain the proton spin puzzle revealed by polarized
deep--inelastic scattering experiments.
In particular, models with {\it negatively} polarized $\bar{s}s$ pairs
in the proton wave function predict $D_{nn}<0$, whereas models with
{\it positively} polarized gluons would predict $D_{nn}>0$.
\end{abstract}

\vspace {-0.10in}

\vskip1cm

\vspace{0.2cm}
CERN--TH/95--47

DOE/ER/40427--04--N95

February 1995

\newpage
The reaction $\bar{p}p\to\bar{\Lambda}\Lambda$ is a testing-ground for
different approaches to non-perturba- tive QCD, in particular the quark
model and meson-exchange models. In addition to total cross section and
angular distribution measurements at different energies, the spin
correlation of the $\Lambda$ and $\bar{\Lambda}$ have been measured
\cite{PS185}.
These were found to be predominantly in a spin-triplet state, with the
spin-singlet component very small and consistent with zero within errors.
 This feature could
easily be understood within quark models, if the $\bar{s}s$ pair that
carry the $\bar{\Lambda}$ and $\Lambda$ spins in the na\"{\i}ve constituent
quark model were produced by effective vector $(^3S_1)$ or scalar $(^3P_0)$
field exchange \cite{AHW}.
Spin-triplet dominance could
also be accomodated in a meson-exchange model, if the relative phase of
the $K-$ and $K^*-$ exchange amplitudes was suitably adjusted \cite{MEX}, but
spin-singlet suppression could not be regarded as a natural prediction
of this class of models.
\vskip0.3cm

There is a plentiful evidence from other experiments at LEAR and elsewhere
that baryon wave functions may be more complicated than in the na\"{\i}ve
constituent quark model. In particular, the experimental value of the
$\pi-$nucleon $\sigma-$term \cite{sigma} and deep-inelastic experiments
\cite{DIS} provide evidence
for hidden $\bar{s}s$ pairs in the nucleon.
Most strikingly, several
recent LEAR experiments \cite{LEAR} find clear evidence for apparent violations
of
the OZI quark-line rule in $\bar{p}N\to\phi X$ annihilations, where
$X=\gamma,\pi,\pi\pi$. A natural interpretation of these data is in terms
of the shake-out or rearrangement of $\bar{s}s$ pairs present in the
$\bar{p}N$ initial state \cite{EGK}.
\vskip0.3cm

It recently was pointed out \cite{EKKS} that many features of
these apparently OZI--violating hadronic processes can be understood if
one assumes that the proton wave function
contains an admixture of {\it polarized} $\bar{s}s$ pairs.
This assumption is motivated by the experimental results on deep-inelastic
scattering \cite{DIS} which indicate that strange quarks and antiquarks in
the proton indeed have a net polarization opposite to the proton spin
\cite{EK}. An alternative interpretation of these deep-inelastic results
ascribes them to to polarized gluons in the proton \cite{gluon}, a suggestion
whose
implications for low--energy $\bar{p}p$ annihilation have not yet been
explored.
\vskip0.3cm

The PS185 Collaboration is now proposing \cite{prop} an extension of its
studies
using a polarized target and measuring the depolarization $D_{nn}$
($p\to\Lambda$
polarization transfer). Quark models generally predict positive values
for this quantity \cite{pos1}, whereas meson exchange models
generally predict
negative values \cite{neg}. We argue in this note that these measurements may
discriminate between the polarized $\bar{s}s$ and gluon interpretations
of the experimental results on polarized deep-inelastic scattering.
Specifically, we find that the polarized $\bar{s}s$ model predicts
{\it negative} depolarization $D_{nn}<0$, whereas the polarized gluon model
predicts
{\it positive} depolarization $D_{nn}>0$. Thus the proposed extension of
the PS185 experiment
could provide valuable insight into the proton spin puzzle.
\vskip0.3cm

The mechanism which is responsible for the negative polarization of the
strange sea is most probably of nonperturbative nature. Its origin can
be linked to chiral dynamics
\cite{BEK}, and
we shall discuss now a particular model based on this idea.
We base our discussion on two starting points.
First, the fact that the masses of
 pions and kaons are small at the typical hadronic scale can
be attributed to the existence of strong attraction between quarks
and antiquarks in the
pseudoscalar $J^{PC}=0^{-+}$ channel. Second, from phenomenological
analyses of the strange quark condensate in the framework of the QCD sum rules
\cite{SVZ} it is known
that the density of strange quark-antiquark pairs in QCD vacuum
is quite high \cite{str}:
$ <0|\bar{s}s|0>\simeq (0.8\pm 0.1) <0|\bar{q}q|0>$.
Using the standard value of the light quark condensate \cite{SVZ},
$<0|\bar{q}q|0>\simeq(250 MeV)^3$, we come to the conclusion that the
density of strange quark-antiquark pairs in the vacuum is about
$1\ fm^{-3}$.
\vskip0.3cm
Let us now consider the basic $|uud>$ proton state
immersed in the QCD vacuum. The strong attraction in the spin-singlet
pseudoscalar channel discussed above will induce correlations between
 light valence quarks from the proton
wave function and vacuum strange antiquarks with opposite spins (see Fig.1).
As a consequence of this,
the spin of the
strange antiquarks will be aligned {\it opposite}
to the proton spin. Moreover, we note that in order to preserve the vacuum
quantum numbers ($J^{PC}=0^{++}$),
strange quark-antiquark pairs must be in a relative spin-triplet,
$L=1$ $^3P_0$ state (see Fig.1).
Therefore the spin of strange quarks must also be aligned {\it opposite} to the
proton spin. The resulting wave function of the $\bar{s}s$ containing
component, consistent with parity and spin constraints, corresponds
to a spin-triplet, polarized $S_z=-1$ $\bar{s}s$ pair with
angular momentum $L_z=+1$ coupled to the ``usual"
$S_z=1/2$ $|uud>$ state. This wave
function is similar to the one used in \cite{EKKS}, which makes identical
predictions for triplet--dominance and depolarization in
 $\bar{p}p\to\bar{\Lambda}\Lambda$.
\vskip0.3cm

The picture advocated above should be contrasted with a similar but
inequivalent one based on effective chiral theories with direct
quark--Goldstone couplings \cite{Gold}:
\begin{equation}
\L_{int} \sim \bar{\Psi}\gamma^{\mu}\gamma_5\Psi \partial_{\mu}\varphi,
\label{L}
\end{equation}
where $\Psi$ is a quark field, and $\varphi$ is the field of a (pseudoscalar)
Goldstone boson.
In these theories, a light quark can emit a spin-zero Goldstone boson, and
this induces spin-flip of the quark. If the emitted boson is a K-meson,
the emission turns the light quark into a strange quark with the opposite
spin orientation (see Fig.2a). As before, this leads to the polarization of
strange quarks
opposite to the spin of the proton. The K-meson can in turn dissociate into
a strange antiquark and light quark, which leads to formation of the
$|uud\bar{s}s>$ component
considered above (see Fig.2b). However the Goldstone fields now are
to be treated as elementary, spin-zero fields, and as such they dissociate
into an {\it unpolarized} $(q\bar{s})$ pair. Though the net
polarization carried by the $\bar{s}s$ pair is again opposite to the proton
spin, the $\bar{s}s$ pair itself can be in either a spin-triplet or
spin-singlet state with statistical weights.
\vskip0.3cm

We observe that the mechanism responsible for the
contribution of the strange sea to the proton spin can be tested
in the process of proton--antiproton annihilation into the hyperon--
antihyperon pair, $\bar{p}p\to\bar{\Lambda}\Lambda$. In the model we advocate
above,
this process can be viewed \cite{EKKS} as the dissociation of a spin-triplet
$\bar{s}s$ pair from the initial proton or antiproton into a
$\bar{\Lambda}\Lambda$ state (see Fig.3).
Since the spin of the $\Lambda$ is carried by the spin of
the strange quark,  this (spin correlation conserving) dissociation
leads to a spin--triplet final state for the two hyperons\footnote{Studies
of initial-- and final--state interactions suggest that these do not affect
significantly the simple polarization arguments we present here and elsewhere
in this paper. Calculations for several different initial- and final-
state interactions show changes in $D_{nn}$ of order $30\%$, averaged over
scattering angle. However, the distinction in sign between the quark models
and meson-exchange models persists, i.e. $D_{nn}$ is positive for quark
models and negative for
meson-exchange models \cite{pos1},\cite{neg}.}. This is indeed
consistent with the experimental observation \cite{PS185} that the
spin--singlet fraction in the $\bar{\Lambda}\Lambda$ final state is
equal to zero within statistical errors. On the other hand, the effective
chiral theory (\ref{L}) would not lead {\it a priori} to this conclusion.
\vskip0.3cm

However, dominance by the spin--triplet state of $\bar{\Lambda}\Lambda$
is not unique to the ``intrinsic strangeness" model. It is also
a feature of the na\"{\i}ve quark model approach to this process \cite{AHW},
since in this approach the $\bar{s}s$ pair is produced through the
$\bar{q}q\to\bar{s}s$ subprocess mediated either by a gluon exchange or by
an effective scalar field; in both cases the structure of the $\bar{s}s$
producing vertex ($^3S_1$ and $^3P_0$ respectively) allows only a spin--triplet
$\bar{\Lambda}\Lambda$ final state \cite{AHW}.
\vskip0.3cm

There exists, however, a way to test the polarized intrinsic
strangeness model in the $\bar{p}p\to\bar{\Lambda}\Lambda$ process.
This model predicts more than just the dominance of the spin-triplet
state in $\bar{\Lambda}\Lambda$. Since the initial $\bar{s}s$ pair
carries a polarization opposite to the (anti)proton spin, it predicts
that the spin of the final $S=1$ $\bar{\Lambda}\Lambda$ pair is polarized
in the direction opposite to the spin of initial spin-triplet $\bar{p}p$
state.
\vskip0.3cm

An experimental observable \cite{prop} which measures the amount of
spin transferred
from the initial-state proton to the final-state hyperon
is the depolarization $D_{nn}$. Assuming a fully polarized proton target,
the depolarization $D_{nn}$ ($\vec{n}$ is normal to the production plane)
is $+1$ if the spin of the final-state $\Lambda$ hyperon is always parallel
to the spin of the target, and $-1$ if the spin of the $\Lambda$ is
always opposite to the spin direction of the target.
The polarized intrinsic strangeness model in the idealized version described
above therefore predicts $D_{nn}=-1$.
\vskip0.3cm

We contrast this prediction with what we would expect within the
polarized gluon interpretation of the proton spin puzzle. According to
one favoured formulation of this interpretation \cite{gluon},
the negative experimental
value of the axial current matrix element
\begin{equation}
<p|\bar{s}\gamma_{\mu}\gamma_{5}s|p> = \Delta s\ s_{\mu}, \label{spin}
\end{equation}
where $s_{\mu}$ is the proton spin vector, is due to the $U(1)$ axial
anomaly, which induces a correction:
\begin{equation}
\Delta s = \Delta \hat{s} - {\alpha_{s} \over {2\pi}}\ \Delta G,
\end{equation}
where $\Delta \hat{s}$ is the polarized $\bar{s}s$ contribution prior
to quantum corrections, and $\Delta G$ is the net gluon polarization in
the proton. In the most absolute version of this interpretation,
$\Delta \hat{s}$ could vanish and the {\it negative} measured value
$\Delta s <0$ could be entirely due to a {\it positive} value of
$\Delta G$. If intrinsic gluons were responsible for $\bar{s}s$ production
during the $\bar{p}p$ annihilation via the
perturbative vertex
\begin{equation}
\L_{QCD} \sim \bar{\Psi}\gamma_{\mu}\Psi\  G^{\mu}, \label{QCD}
\end{equation}
the $\bar{s}s$ pair would be produced in a spin-triplet state, as
inferred from the $\bar{\Lambda}\Lambda$ spin correlations.
{\it However}, if $\Delta G > 0$ as suggested in the gluon interpretation
of the proton spin puzzle, the depolarization should be {\it positive}:
$D_{nn}>0$ (see Fig.4). We note however that the more conventional quark model
based on effective vector exchange also predicts positive
depolarization \cite{pos1}.
\vskip0.3cm
Thus the measurement of the depolarization in the
$\bar{\Lambda}\Lambda$ process could serve as an interesting test of the
dynamics responsible for the apparently ``anomalous" decomposition of
the proton spin.
\vskip0.5cm
We thank M. Karliner and M.G. Sapozhnikov for enjoyable collaboration at the
initial stage of this work, E.M. Henley and L. Wilets for useful discussions,
and K. R\"ohrich for his interest in this work and stimulating suggestions.

The work of M.A. was supported in part by the U.S. Department of Energy and
by the National Science Foundation, grant no. PHY-9223618. D.K. acknowledges
support of the German Research Ministry (BMFT) under the Contract 06 BI 721.

\newpage
\medskip
{\large\bf Figure captions}
\vskip0.3cm

Fig.1 a,b) Strong correlation between light valence quarks  and vacuum
strange antiquarks in the spin--singlet pseudoscalar channel induces a
spin--triplet
$\bar{s}s$ component of the proton wave function aligned opposite to the
proton spin.  (In all figures the direction of spin quantization is taken
normal to the plane of the quark line diagrams.)
\vskip0.3cm

Fig.2 a) The emission of a $K^+$ meson turns the light quark into a strange
quark with the opposite spin orientation.
b) Dissociation of the $K$ meson leads to formation of an $\bar{s}s$
component with {\it net} polarization opposite to the proton spin,
but the $\bar{s}s$ pair can be in either a spin--triplet or a spin--singlet
state with {\it a priori} statistical weights.
\vskip0.3cm

Fig.3 The $\bar{p}p\to\bar{\Lambda}\Lambda$ process viewed as the dissociation
of a spin--triplet $\bar{s}s$ pair from the initial state proton
(or antiproton) wave function
into a $\bar{\Lambda}\Lambda$ state. The spin of the produced $\Lambda$
is always opposite to the spin of the initial proton.

\vskip0.3cm

Fig.4 The $\bar{p}p\to\bar{\Lambda}\Lambda$ process viewed as the dissociation
of a polarized gluon from the initial state proton (or antiproton) wave
function
into a spin--triplet $\bar{s}s$ state. The spin of the produced $\Lambda$
is parallel to the spin of the gluon, which is in turn parallel to the spin
of the initial proton.

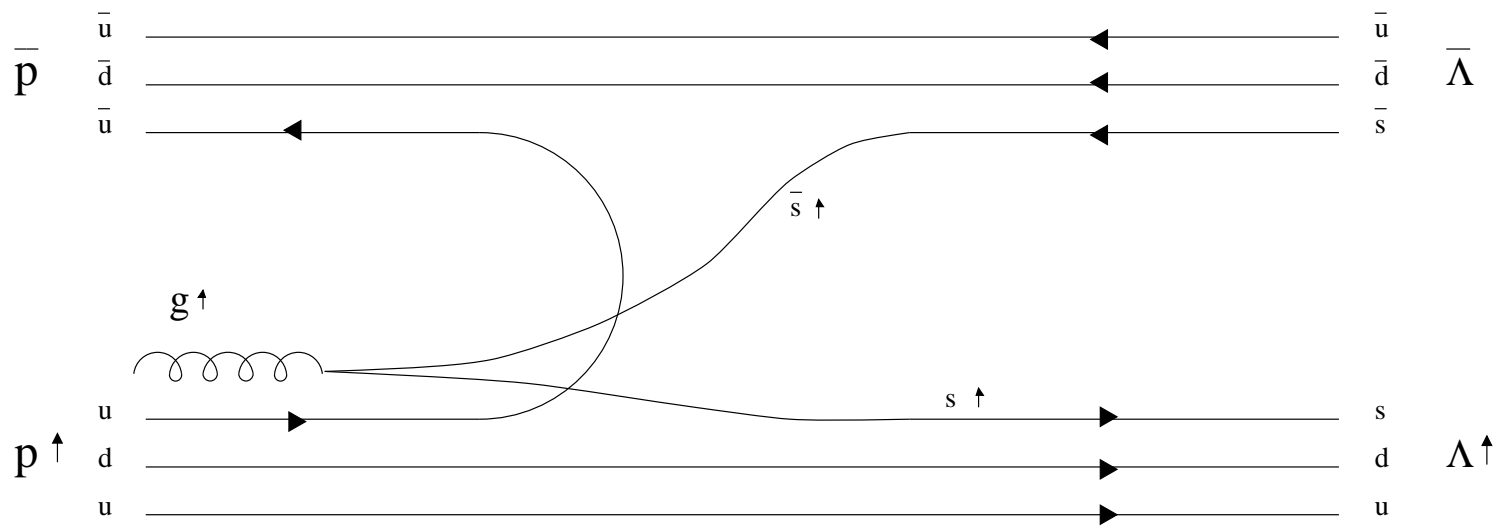
\end{document}